\documentclass[twocolumn,showpacs,prb]{revtex4}
\usepackage{bm}
\usepackage{graphicx}

\usepackage{amsmath}
\usepackage{amsfonts}
\newcommand{\nix}[1]{}

\begin{document}

\title{Spin-dependent recombination in GaAsN alloys}
\author{V.K.~Kalevich}
\author{E.L.~Ivchenko}
\author{M.M.~Afanasiev}
\author{A.Yu.~Egorov}
\author{A.Yu.~Shiryaev}
\author{V.M.~Ustinov}
\affiliation{A.F.~Ioffe Physico-Technical Institute, St.
Petersburg 194021, Russia}
\author{B.~Pal}
\author{Y.~Masumoto}
\affiliation{Institute of Physics, University of Tsukuba, Tsukuba
305-8571, Japan}
\begin{abstract}
The spin-dependent recombination (SDR) has been observed in
GaAs$_{1-x}$N$_x$ ($x$ = 2.1, 2.7, 3.4\%) at room temperature. It
reveals itself in a decrease of the edge photoluminescence (PL)
intensity by more than a factor of 3 when either the polarization
of the exciting light is changed from circular to linear or the
transverse magnetic field of $\sim$300 gauss (G) is applied. The
interband absorption of the circularly polarized light results in
a spin polarization of conduction electrons, which reaches 35\%
with increasing the pump intensity. The effects observed are
explained by dynamical polarization of deep paramagnetic centers
and spin-dependent capture of conduction electrons by these
centers. The PL depolarization in a transverse magnetic field
(Hanle effect) allows us to estimate the electron spin relaxation
time in the range of 1 ns. Theoretically, it has been concluded
that, due to the SDR, this long time is controlled by slow spin
relaxation of bound electrons. In all the three alloy samples, a
positive sign of the bound-electron $g$-factor is determined
experimentally from the direction of their mean spin rotation in
the magnetic field.
\end{abstract}


\maketitle In recent years Ga(In)AsN alloys grown on GaAs
substrates have attracted an increasing interest due to their
unusual physical properties and application in near-infrared
optoelectronics. The interaction of nitrogen localized
isoelectronic states with the conduction-band extended states
leads to an anomalous reduction in the band gap for a small, a few
\%, nitrogen content~\cite{Kondow,Waluk}. Evidently, the
understanding of the free-carrier kinetics and recombination
mechanisms in these materials are important from the point of view
of both fundamental physics and practical applications. In the
present work the SDR has been demonstrated in GaAsN alloys for the
first time. It manifests itself at room temperature in a strong
decrease of the edge PL intensity when the polarization of
exciting light is changed either from circular to linear or the
transverse magnetic field of $\sim$300 G is switched on. A high
spin polarization ($\sim$35\%) and long-term spin memory ($\sim$1
ns) of conduction electrons, both related to the SDR, were
observed also at room temperature. We studied the 0.1 $\mu$m thick
nominally undoped GaAsN films grown by the MBE on a
semi-insulating (001) GaAs substrate between GaAs
layers~\cite{Egorov}. Arsine was used as the source of arsenic.
After the growth, the structures were annealed for 5 minutes at
700$^{\circ}$C in the flow of arsenic in the growth chamber.
Investigated were 3 samples with nitrogen content $x$ = 2.1, 2.7,
3.4\%.

The spin polarization of conduction electrons was created under
interband absorption of the circularly polarized light~\cite{OO}.
Its magnitude was determined by measuring the degree of PL
circular polarization $\rho = (I^+ - I^-)/(I^+ + I^-)$, where
$I^+$ and $I^-$ are the intensities of the right ($\sigma^+$) and
left ($\sigma^-$) circularly polarized PL components,
respectively. A highly sensitive polarization
analyzer~\cite{Kulkov} with an InGaAsP photomultiplier, a quartz
polarization modulator, and a two-channel photon counter
synchronized with the polarization modulator were used to measure
$\rho$ and the total intensity $I = I^+ + I^-$ in a spectral range
up to 1.4 $\mu$m. PL was excited by tunable Ti:saphire laser along
the normal to the sample and recorded in the backscattering
geometry along the growth axis $z
\parallel [001]$. The measurements were carried out at 300 K. As
the main experimental results are qualitatively the same for all
the samples under study, we present below the data for
Ga$_{1-x}$N$_x$ alloy with $x$ = 2.1\%.

Fig.~1 shows the PL spectra and circular polarization measured in
GaAs$_{0.979}$N$_{0.021}$ sample under excitation by circularly
polarized ($\sigma^+$ or $\sigma^-$) light of a high intensity (an
increase in $x$ in the samples with $x$ = 2.7\% and $x$ = 3.4\%
brings about a strong red shift of the spectra). Each PL spectrum
consists of two strongly overlapping bands, their splitting grows
with the increasing nitrogen content and reaches 50 meV at $x$ =
3.4\%~\cite{Egorov}. Relative to the polarization of the pump
beam, the low- and high-energy bands are polarized, respectively,
negatively and positively. The presence of two PL bands and the
opposite signs of their polarization were interpreted in our
previous works~\cite{Egorov,Nano} as originating from the
strain-induced splitting of the light- and heavy-hole subbands of
the complicated valence band $\Gamma_8$. Indeed, the smaller
lattice constant of GaAsN as compared with GaAs results in a
biaxial tension of the GaAsN film in the interface plane, which is
equivalent to a uniaxial compression along the growth axis. The
uniaxial strain is accompanied by splitting of light- and
heavy-hole subbands, which increases with increasing $x$. The
polarization spectrum in Fig.~1 was measured under simultaneous
excitation of electrons from the both valence subbands. In order
to exclude an excitation of electrons into GaAs barrier, the
energy of exciting quanta was smaller than the band gap of GaAs.
Thus, the negative polarization of the low-energy PL band and the
positive polarization of the high-energy PL band allow one to
conclude that those bands arise due to the radiative recombination
of electrons with light and heavy holes, respectively.

\begin{figure}[t]
  \centering
    \includegraphics[width=.47\textwidth]{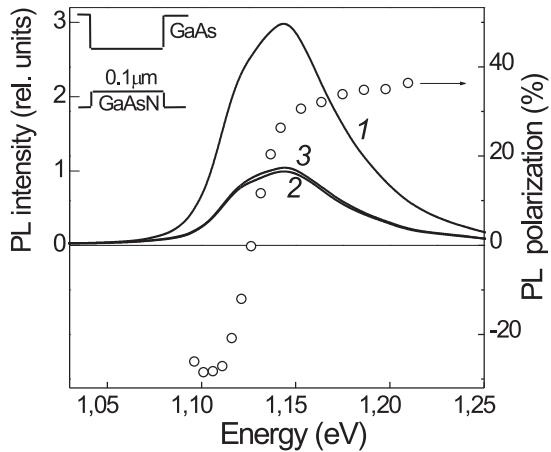}
\caption{PL spectra under excitation by circularly (curves 1, 3)
and linearly (curve 2) polarized light and the PL circular
polarization spectrum (circles) of GaAs$_{0.979}$N$_{0.021}$
layer. Spectrum 3 is obtained in the transverse magnetic field
$B=$ 400 G. Excitation energy $h \nu_{\rm exc}$ = 1.305 eV;
excitation intensity $J = 150$ mW; T = 300 K.}
\end{figure}

Spectrum 2 in Fig.~1 is obtained under excitation by linearly
($\pi$) polarized light. It is seen from curves 1 and 2 that the
change of excitation light polarization from circular to linear
leads to a decrease of PL intensity $I$ near the PL maximum by a
factor of three. The same decrease in $I$ is found when the
transverse magnetic field of $\sim$300 G is switched on. In
comparison with these changes, the variation of $I$ observed
following a change of $\sigma^+$ to $\sigma^-$ excitation is
negligibly small. Fig.~2a represents the dependence $\rho(B)$
measured  in GaAs$_{0.979}$N$_{0.021}$ in a transverse magnetic
field ${\bf B} \perp [001]$ at different intensities of the
$\sigma$ excitation near the PL-band maximum involving heavy
holes. The dependencies are well approximated by Lorentz curves
(solid lines in Fig.~2a) of the form
\begin{equation} \label{rb}
\rho(B) = \rho^* + \frac{\rho_0}{1 + (B/B_{1/2})^2}\:\:\:,
\end{equation}
where $\rho^*$ is a constant component ($\approx$5\%), $B_{1/2}$
is the half-width of the curve at half-height, $\rho_0$ is the
maximum change of $\rho$ in the magnetic field. The values of
$B_{1/2}$ and $\rho_0$ are strongly dependent on the pump
intensity. Additional measurements show that the widths of the
curves $\rho(B)$ are equal for the positive ($c$-$hh$ transition)
and the negative ($c$-$lh$ transition) polarization of PL measured
with equal intensity of the pump light. This confirms the
assumption that the photoholes are spin-unpolarized because of
their fast spin relaxation and, hence, the PL polarization is
caused by the polarization of electrons solely.

Fig.~2b shows the magnetic-field dependencies $I(B)$. They can
also be fitted by Lorentz curves (solid curves) with the widths
coinciding with those of the Hanle curves $\rho(B)$ taken at the
same excitation intensity.

Experimental results presented in Figs.~1 and 2 can be explained
in the SDR model proposed by Weisbuch and Lampel~\cite{Weisbuch}
for Ga$_{0.6}$Al$_{0.4}$As solid solutions and applied in
Refs.~[\onlinecite{Miller,Paget}] while analyzing the
recombination processes in GaAs crystals è GaAs/AlGaAs multiple
quantum-well structures. We use this model and assume that each
deep center can contain either one electron with noncompensated
spin $\pm 1/2$ or two electrons in the singlet state with the zero
total spin. Under normal incidence of the circularly polarized
light in the absence of a magnetic field, the electronic spins are
polarized along the excitation direction (the axis $z$). Let us
introduce the notations $n_{\pm}, N_{\pm}$, respectively, for the
concentration of the free electrons or paramagnetic centers with
the component $\pm 1/2$ of the electron spin, and $N_{\uparrow
\downarrow }$ for the concentration of centers with two electrons.
In the SDR model under consideration the contribution of the
band-to-band recombination is neglected in the balance equations
and the rate of free-carrier capture by deep levels is described
by $\bigl( d n_{\pm}/dt \bigr)_{\rm rec} = - \gamma_e n_{\pm}
N_{\mp}$ for electrons and $\bigl( d p/dt \bigr)_{\rm rec} = -
\gamma_h p N_{\uparrow \downarrow}$ for holes, where $p$ is the
density of holes assumed to be unpolarized irrespective of the
incident-light polarization. In addition to the recombination
constants $\gamma_e$ and $\gamma_h$, the system is characterized
by two times, namely, by the spin relaxation times of free and
bound electrons, respectively, $\tau_s$ and $\tau_{sc}$.
Apparently, at room temperature $\tau_s \ll \tau_{sc}$. As a
result, the rate equations for $n_{\pm}$ and $N_{\pm}$ take the
form
\begin{eqnarray} \label{1}
\frac{dn_{\pm}}{dt} &=& - \gamma_e n_{\pm} N_{\mp} - \frac{n_{\pm}
- n_{\mp}}{2 \tau_s} + G_{\pm}\:,\\
\frac{dN_{\pm}}{dt} &=& - \gamma_e n_{\mp} N_{\pm} -
\frac{N_{\pm} - N_{\mp}}{2 \tau_{sc}} + \frac12 \gamma_h p
N_{{\uparrow \downarrow}} \:,\nonumber
\end{eqnarray}
where $G_{\pm}$ are the photogeneration rate of electrons with the
spin $\pm 1/2$.

\begin{figure}[t]
  \centering
    \includegraphics[width=.45\textwidth]{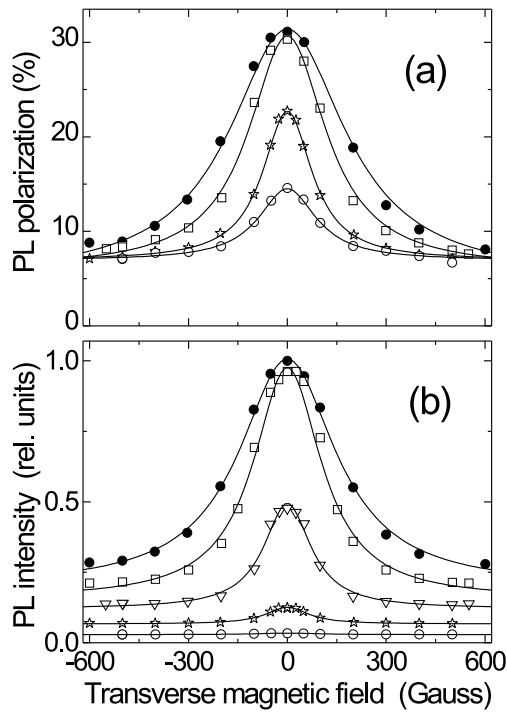}
  \caption{Degree of PL circular polarization (a) and PL intensity
(b) in the GaAs$_{0.979}$N$_{0.021}$ layer vs. transverse magnetic
field at different intensities $J$ of the circularly polarized
excitation. Excitation energy $h \nu_{\rm exc} = 1.311$ eV;
detection energy $h \nu_{\rm det} = 1.159$ eV; $T$ = 300 K.
$J$(mW) = 240 ($\bullet$), 150 ($\square$), 100 ($\nabla$), 50
($\star$), 25 ($\circ$). Solid lines are the best fit of the
experimental data by the Lorentz curves $y(B) = y^* + y_0(1 +
B^2/B^2_{1/2})^{-1}$.}
\end{figure}

For a qualitative interpretation of the behavior of the PL
intensity and circular polarization as functions of the
incident-light polarization and the magnetic field strength, it
suffices to analyze the limiting case of weak photoexcitation
where $P_c$ is small and one has~\cite{Weisbuch,Paget}:
\begin{equation} \label{j1}
I \propto 1 + a P_c P\hspace{2 mm},\: \hspace{5 mm}\rho = b (P_i +
P_c)\:.
\end{equation}
Here $P= (n_+ - n_-)/(n_+ + n_-)$ and $P_c= (N_+ - N_-)/(N_+ +
N_-)$ are the degrees of spin polarization of free and bound
electrons, $P_i = (G_+ - G_-)/(G_+ + G_-)$ is the
spin-polarization degree of photoelectrons in the moment of
generation, and $a, b$ are positive coefficients. Since the
linearly polarized excitation gives rise to no optical orientation
of electron spins, the ratio $I({\rm circ})/I({\rm lin})$ of PL
intensities under circularly and linearly polarized pumping is
given by $(1 + a P_c P)$, where values of $P_c, P$ refer to
circularly polarized excitation. Note that the product $P_cP$ is
positive and independent of the sign of the incident circular
polarization. Therefore, $I({\rm circ}) > I({\rm lin})$. In a
transverse magnetic field, ${\bf B} \perp z$, so that $|g_e|\mu_B
B \tau_s /\hbar \ll 1$ but values of $|g_c|\mu_B B \tau_{sc}
/\hbar$ are arbitrary ($g_e, g_c$ are the free- and bound-electron
$g$-factors, $\mu_B$ is the Bohr magneton), one can neglect the
rotation of free-electron spin due to the Larmor precession. In
this case Eq.~(\ref{j1}) is valid for non-zero $B$ as well if
$P_c$ is understood as the degree of bound-electron spin
polarization along the $z$ axis, $P_c(B) = P_c(0)[1 + (\Omega_c
T_{sc})^2]^{-1}$. Here $\Omega_c = g_c \mu_B B/\hbar$ is the
bound-electron Larmor frequency, $T_{sc}^{-1} = \tau^{-1}_{sc} +
\tau_c^{-1}$, $\tau_c^{-1} = \gamma_h N_{\uparrow
\downarrow}(p/N)$, because the depolarization of bound-electron
spins occurs both as a result of the spin relaxation (determined
by the time $\tau_{sc}$) and due to the capture of holes by the
centers with two electrons (determined by the time $\tau_c$). The
increasing magnetic field results in a depolarization of the
bound-electron spins reducing the intensity from $I({\rm circ})$
to $I({\rm lin})$ and the polarization $\rho$ from $b (P_i + P_c)$
to $b P_i \equiv \rho^*$.

In order to find the PL intensity and polarization in case of
strong pumping one needs to solve the set of nonlinear rate
equations. Preliminary calculation shows that the above SDR model
is applicable for the explanation of experimental data obtained in
the whole investigated range of $J$.

Here we show, in the system under study, in spite of the short
spin-relaxation time $\tau_s$, there is also a long time of
evolution of free-electron spins. For this purpose we reduced the
rate equations for $n_{\pm}$ and $N_{\pm}$ to the following
equations for the polarization degrees:
\begin{eqnarray} \label{7b}
\frac{dP}{dt} &=& - \frac{P}{\tau_s} + \frac{P_c}{\tau_0} ( 1 -
P^2) + \frac{G}{n}(P_i - P)
\:, \\
\frac{dP_c}{dt} &=& - \frac{P_c}{T_{sc}} +
\frac{n}{N}\frac{1}{\tau_0} P(1 - P_c^2) \:,
\end{eqnarray}
where $G = G_+ + G_-$ is the total generation rate, and $\tau_0 =
(2/\gamma_e N)$ is the capture time of conduction-band electrons
in the absence of SDR. The second term in the right-hand side of
Eq.~(\ref{7b}) is related to SDR and plays the role of
spin-generation rate. By using Eq.~(\ref{7b}) one can readily show
that, under abrupt cutting off the steady-state optical excitation
$(G=0)$, the polarization time decay $P(t)$ comprises both fast
and slow components: the first one decays within the time
$\sim$$\tau_s$, whereas the second contribution can be
approximated by $P(t) \approx (\tau_s/\tau_0) P_c(t)$.

The proximity of the maximum value of PL polarization ($\sim$35\%)
in Fig.~2a to its limiting value of 50\% indicates a long-term
spin memory of conduction electrons. As shown above, the spin
memory of the nonlinearly coupled spin subsystems of free and
bound electrons is controlled by the long spin lifetime $T_{sc}$
of bound electrons. The latter can be easily determined from the
Hanle effect as $T_{sc} = \hbar/(g_c \mu_B B_{1/2})$, so that the
longer $T_{sc}$ corresponds to the smaller $B_{1/2}$. The smallest
half-width found from Fig.~2a is $B_{1/2}^{\rm min} = (91 \pm 5)$
G. Taking into account that $\tau^{-1}_{sc} < T_{sc}^{-1}$, we
obtain a lower estimate for the spin relaxation time multiplied by
the $g$-factor: $g_c \tau_{sc} > g_c T^{\rm min}_{sc} = (1.3 \pm
0.2)$ ns. The magnitude of $g_c$ in GaAsN is unknown (the
available study of the electron $g$-factor in nitrogen-containing
GaAs-based alloys is devoted to measurements of $g$ in
GaInAsN~\cite{Skier}). In a special experiment with oblique
incidence of light onto the sample, when the PL is measured in the
backscattering geometry at an angle to the pump beam, and the
magnetic field is normal to the directions of excitation and
observation~\cite{Kalevich} we found that $g_c > 0$. Similarly,
the positive sign of $g_c$ was found in GaAsN films with the 2.7
and 3.4\% nitrogen content. The sign of $g_c$ allows us to make up
a conclusion that its value is close to the $g$-factor of a free
electron in vacuum, i.e., $g_c \approx 2$. It follows then that
the bound-electron spin relaxation time $\tau_{sc}$ exceeds 0.6
ns.

In conclusion, the strong SDR has been observed in GaAsN alloys at
room temperature. Giant electron spin lifetime and high
polarization have been found under optical pumping conditions.
Therefore, the nitrogen-containing alloys GaAsN can be considered
as a promising material for the development of spintronic devices.
The sensitivity of the PL circular polarization and the Hanle
effect half-width to the pump intensity has been firstly observed
in a semiconductor of intrinsic conductivity type. A simple model
of SDR has been applied to explain the observed dependencies of
the electron lifetime, the electron spin lifetime and the degree
of electron spin orientation on the pump intensity. It has been
shown that, in spite of the very short spin relaxation time of
free electrons, the behavior of the coupled system of
spin-polarized free and bound electrons is controlled by the long
spin relaxation time of bound electrons that amounts 1 ns. Using
the Hanle effect, we have determined the positive sign of
$g$-factor of bound electrons which are responsible for the
observed SDR.

The work is carried out with partial support of the Russian
Foundation for Basic Research and the Japan Society for the
Promotion of Science. The authors are grateful to K.V.~Kavokin and
N.V.~Kryzhanovskaya for fruitful discussions.

\end{document}